\documentclass[twocolumn,a4paper,amsmath,amssymb]{revtex4}
\usepackage{graphicx}
\usepackage{times}
\usepackage{amsmath}
\usepackage{braket}
\usepackage{xcolor}
\usepackage{natbib}
\usepackage{dsfont}

\renewcommand{\eqref}[1]{Eq.~(\ref{#1})}

\newcommand{\figref}[1]{Fig.~\ref{#1}}

\newcommand{\removedD}[1]{{\color{gray}{#1}}}
\renewcommand{\removedD}[1]{{}} 
\newcommand{\CE}[1]{{\color{red}{}}}
\newcommand{\ce}[1]{{{#1}}}

\renewcommand{\eqref}[1]{Eq.~(\ref{#1})}

\newcommand{\secref}[1]{Section~\ref{#1}}

\newcommand{\appref}[1]{\hyperref[#1]{Appendix~\ref*{#1}}}
\newcommand{\tabref}[1]{\hyperref[#1]{Table~\ref*{#1}}}

\begin{document}
\title{Controlling the dynamic range of a Josephson parametric amplifier}
\author{C. Eichler and A. Wallraff}
\affiliation{Department of Physics, ETH Z\"urich, CH-8093, Z\"urich, Switzerland.}
\date{\today}
\begin{abstract}
One of the central challenges in the development of parametric amplifiers is the control of the dynamic range relative to its gain and bandwidth, which typically limits quantum limited amplification to signals which contain only a few photons per inverse bandwidth. Here, we discuss the control of the dynamic range of Josephson parametric amplifiers by using Josephson junction arrays. We discuss gain, bandwidth, noise, and dynamic range properties of both a transmission line and a lumped element based parametric amplifier. Based on these investigations we derive useful design criteria, which  may find broad application in the development of practical parametric amplifiers.
\end{abstract}
\maketitle
\section{Introduction}
\label{ch:paramp}
Due to the rapidly evolving field of \ce{quantum optics and information processing} with superconducting circuits the interest in low-noise amplifiers has dramatically increased in the past five years and has lead to a body of dedicated research on Josephson junction based amplifiers \cite{Yurke2006,Castellanos2007,Tholen2007,Kinion2008,Yamamoto2008,Castellanos2008,Palacios-Laloy2008,Kamal2009,Bergeal2010,Hatridge2011,Vijay2011,Gao2011,Eichler2011a,Abdo2012,Kamal2012}.
The most successful quantum limited detectors which have so far been realized in the microwave frequency range are based on the principle of parametric amplification \cite{Louisell1961,Gordon1963,Mollow1967,Clerk2010}.
Josephson parametric amplifiers (JPAs) have not only been used to generate squeezed radiation \cite{Castellanos2008,Mallet2011,Eichler2011a,Flurin2012,Menzel2012}, but moreover enabled the realization of quantum feedback and post-selection based experiments \cite{Vijay2012,Johnson2012,Riste2012,Campagne-Ibarcq2013}, the efficient displacement measurement of nanomechanical oscillators \cite{Teufel2011} and the exploration of higher order photon field correlations \cite{Eichler2012,Eichler2012b}.

While JPAs have been demonstrated to operate close to the quantum limit, their performance is to date mostly limited by their relatively small dynamic range. Here, we discuss the control of the dynamic range by making use of Josephson junctions arrays in the parametric amplifier circuit, which we have already employed in recent experiments \cite{Eichler2012b,Steffen2013}. After reviewing the principles of parametric amplification we discuss  bandwidth and noise constraints in dependence on the circuit design, based on which we derive simple strategies for optimized circuit design.
\section{Principles of parametric amplification}
\subsection{Parametric processes at microwave frequencies}
\begin{figure*}[t]
\centering
\includegraphics[scale=1]{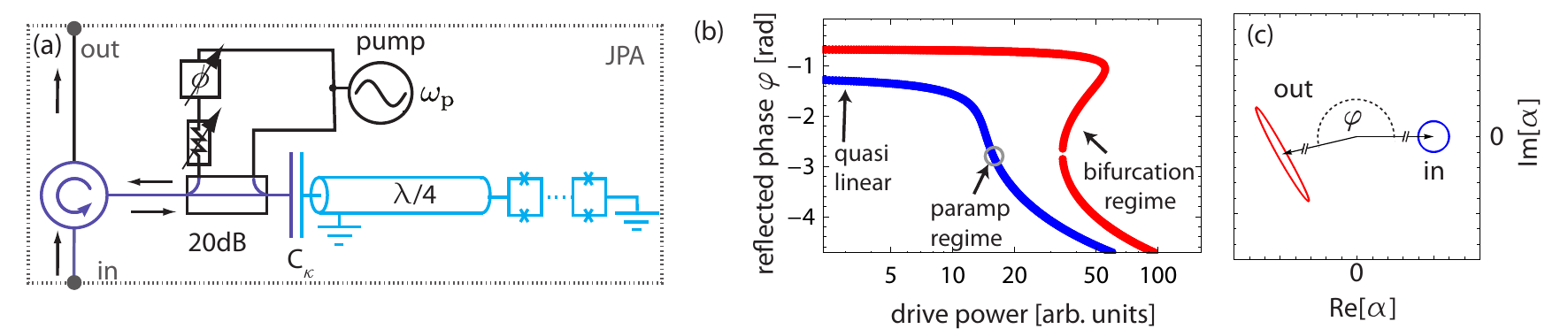}
\caption{(a) Circuit diagram of a transmission line resonator based parametric amplifier. The resonator is coupled with capacitance $C_\kappa$ to a transmission line where input and output modes are spatially separated using a circulator.  A $20\,{\rm dB}$  directional coupler between the $\lambda$/4-resonator and the circulator is used to apply the pump field required for modulating the SQUID inductance. The second port of the directional coupler can be used to interferometrically cancel out the pump tone reflected from the sample. (b) Phase of the reflected probe signal vs. drive power for two characteristic drive frequencies below (blue) and above (red). (c) Illustration of the nonlinear oscillator response in the quadrature plane. The blue circle represents various input fields $\alpha_{\rm in}$ close to the one indicated by the gray circle in (a). Due to the nonlinear response of the resonator they are transformed into output fields $\alpha_{\rm out}$ indicated by the red ellipse.
}
\label{fig:ParampSchematic}
\end{figure*}
In quantum optics the \ce{word} \emph{parametric} is used  for processes in which a nonlinear refractive medium is employed for mixing different frequency components of light. Such processes are parametric in the sense that a coherent pump field, applied to a nonlinear medium, modulates its refractive index, which appears as a \emph{parameter} in a semi-classical treatement. This time-varying parameter is affecting modes with \ce{frequencies detuned from the frequency of the pump field and can stimulate their population with photons}. The energy for creating these photons is provided by the pump field.

\CE{Predominantly, the frequency conversion is either realized as a three-wave mixing process in media with $\chi^{(2)}$ nonlinearity or as a four-wave mixing process in media with $\chi^{(3)}$ nonlinearity. In a three-wave mixing process one pump photon with frequency $\omega_p$ is converted into a pair of signal $\omega_s$ and idler $\omega_i$ photons obeying energy conservation $\omega_p = \omega_i + \omega_s$. Four-wave mixing describes the conversion of two  pump photons into a pair of signal and idler photons $2\omega_p =\omega_s + \omega_i$. If the signal and idler modes are initially in the vacuum state such processes are called  spontaneous parametric down-conversion and spontaneous four wave mixing \cite{Burnham1970}. This phenomenon is abundantly used for creating heralded single photons, i.e. single photons conditioned on a photon counting event.

In order to enhance the production rate of signal and idler pairs one can place the nonlinear medium inside a cavity to form an optical parametric oscillator \cite{Slusher1985}. The emitted  signal and idler photons are perfectly correlated which results in squeezing. Since the conversion from pump photons into signal and idler pairs is stimulated by already existing fields the described phenomenon is an amplifying process, which is called parametric amplification. }

The refractive index in optics is equivalent to the impedance of electrical circuits. In order to realize parametric processes at microwave frequencies we therefore modulate an effective impedance. This is achieved by  varying \ce{the parameters of} either a capacitive or an inductive element in time. Although there have been early proposals for fast time-varying capacitances \cite{Louisell1960}, it now is considered to be more convenient to make use of dissipationless Josephson junctions for this purpose. In a regime in which the current $I$ flowing through a Josephson junction is much smaller than its critical current $I_C \equiv 2 e E_J/\hbar$ its associated inductance is approximately $L\approx L_J (1+ \frac{1}{6}(I(t)/I_C)^2)$. Applying an $AC$ current through the junction using appropriate microwave drive fields therefore leads to the desired time-varying impedance. Because of the proportionality of the inductance $L$ to the square of the current $I^2(t)$, such a drive results in a four-wave mixing process \cite{Slusher1985}.

The effective impedance can alternatively be modulated by varying the magnetic flux threading a superconducting quantum interference device (SQUID) loop \cite{Yamamoto2008} such that the effective inductance is approximately modulated proportionally to the AC current  $I(t)$ flowing in the loop, $L \approx L_J(1+ I(t)/I_0)$. The quantity $I_0$ in this expression depends on the DC flux bias point of the SQUID loop. Since the relation between current and inductance is in this case linear, the magnetic flux drive results in a three-wave mixing process \cite{Burnham1970}.

In order to enhance parametric amplification in a well-controlled frequency band while suppressing it for frequencies out of this band, the modulated Josephson inductance is frequently integrated into a microwave frequency resonator. This is the simplest way to control the band in which parametric amplification occurs.  A number of variations of this basic idea are now explored. The circuit design has recently been modified to achieve a spatial separation of signal and idler modes \cite{Bergeal2010,Bergeal2010a,Bergeal2012,Roch2012,Flurin2012} and to build traveling wave amplifiers, in which a field is amplified while propagating in forward direction coaxially with a pump field \cite{HoEom2012,Yaakobi2013}. Various drive mechanisms ranging from single and double pumps \cite{Kamal2009} to magnetic flux drives \cite{Yamamoto2008,Wilson2011,Wustmann2013} have been explored. Being aware of this variety of possible approaches, we focus here on a single mode (degenerate) parametric amplifier driven with one pump tone close to its resonance frequency.
\subsection{Circuit QED implementation of a parametric amplifier}
The JPA essentially is a weakly nonlinear oscillator, in which the nonlinearity is provided by Josephon tunnel junctions. In practice, this is typically realized either as a transmission line resonator shunted by a SQUID \cite{Palacios-Laloy2008,Gao2011,Eichler2011a}, see \figref{fig:ParampSchematic}a, or as a lumped element nonlinear oscillator \cite{Vijay2011}. The use of a SQUID instead of single tunnel junction guarantees tunability of the resonance frequency. Since resonator-based parametric amplifiers provide amplification in a narrow band only, tunability is highly desirable to match the band of amplification with the frequency of the signal to be amplified.

The relevant part of the Hamiltonian which describes the parametric amplifier considered here can be written as
\begin{equation}
H_{\rm JPA} = \hbar \tilde{\omega}_{\rm 0} A^\dagger A + \hbar \frac{K}{2} (A^\dagger)^2 A^2,
\label{eq:JPAHamiltonian}
\end{equation}
where $A$ labels the annihilation operator of the intra-resonator field. Expressions for the resonance frequency $\tilde{\omega}_0/2\pi$ and the effective Kerr nonlinearity $K$ are derived in \secref{sec:tunableRes} based on the full circuit model. In the following section we analytically study the dynamics of this system using the input-output formalism. Before presenting the mathematical derivations, we qualitatively describe different dynamical regimes of this nonlinear oscillator and explain the mechanism which leads to amplification.

If we assume for the moment that the JPA has no internal losses, all the incident power is reflected from the resonator and the classical response (i.e. reflection coefficient) is completely specified by the phase $\varphi$ of the reflected field. In contrast to a linear system, where $\varphi$ only depends on the frequency $\omega/2\pi$, it also depends on the  power of the probe field in the case of a nonlinear oscillator. In \figref{fig:ParampSchematic}(b), the theoretically expected value of $\varphi$ is plotted as a function of the probe amplitude for two characteristic drive frequencies. While the phase is constant for low drive powers (quasi-linear response), the phase changes significantly for increased drive power. Depending on the probe frequency we either find a bistable regime where two stable solutions exist \cite{Dykman1980,Marthaler2006} or a regime where the phase  has a unique solution (read and blue data sets in \figref{fig:ParampSchematic}(b)). In both cases the phase significantly depends on the input power.  The bistable response can for example be used to realize a bifurcation amplifier \cite{Siddiqi2004,Vijay2009} and for nonlinear dispersive readout \cite{Mallet2009}, which has been intensely studied in the context of circuit QED.

Since we are particularly interested in \emph{linear} amplification the following discussion is focused on the regime, in which the response has a unique solution (blue data set). The mechanism of amplification can be understood qualitatively in the following way. If we imagine that the device is constantly driven at a frequency and power at which the reflected phase $\varphi$ depends sensitively on power (see gray circle in \figref{fig:ParampSchematic}(b)), the system will strongly react to small perturbations. Such perturbations, which could be caused by an additional small signal field for example, are therefore translated into a large change of the output field.

We illustrate this process leading to amplification by plotting the resonator response for input fields $\alpha_{\rm in}$ with slightly varying amplitude and phase. In \figref{fig:ParampSchematic}(c) we indicate the input fields by a blue circle around the mean value (arrow). The small differences in amplitude of the input field translate into large changes in $\varphi$ of the output field $\alpha_{\rm out}$ (red ellipse). If we interpret the arrow in \figref{fig:ParampSchematic}(c) as a constant pump field and its difference to the points on the blue circle as an additional signal, the signal is either amplified or deamplified depending on its  phase relative to the pump.

The mechanism of amplification can thus be understood intuitively by considering the nonlinear response to a monochromatic drive field. In order to characterize the exact behavior of input fields with finite bandwidth we analyze the response in more detail below.
\section{Input-Output relations for the Parametric amplifier}
\label{sec:parampTheory}
\subsection{Classical nonlinear response}
\label{sec:parampInOut}
\begin{figure}[t]
\centering
\includegraphics[scale=1]{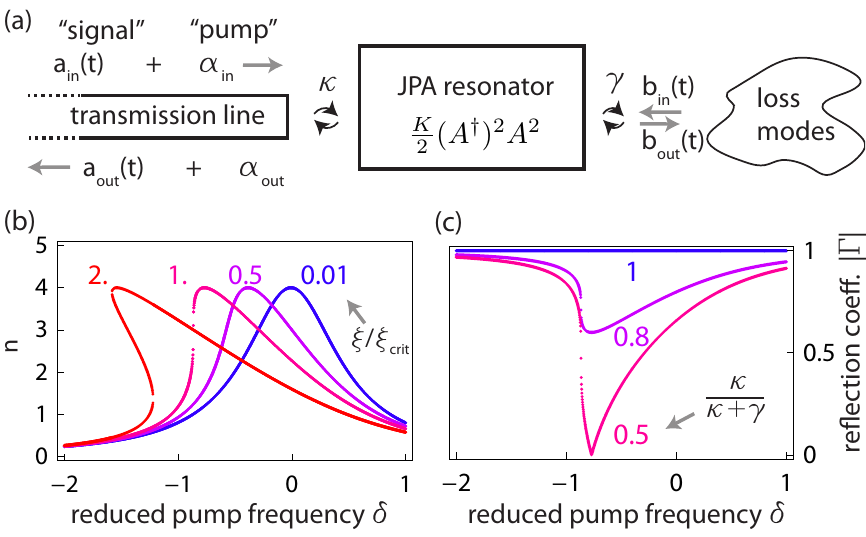}
\caption{(a) Schematic of the input-output model used for calculating the response of the parametric amplifier in the presence of additional loss modes. (b) Normalized pump field photon number $n$ in the resonator as a function of reduced pump frequency $\delta$ for effective drive strengths $\xi/\xi_{\rm crit}=0.01,0.5,1,2$, where $\xi_{\rm crit} = -1/\sqrt{27}$.
(c) Absolut value of the reflection coefficient $|\Gamma|$ for different coupling ratios $\kappa/(\kappa+\gamma)=1,0.8,0.5$. 
}
\label{fig:ParampInputOutput}
\end{figure}

Here, we  employ the input-output formalism \cite{Gardiner1985,Walls1994} to calculate the nonlinear resonator response discussed qualitatively in the previous section. The derivation presented here is inspired by Ref.~\cite{Yurke2006}. A schematic of the input-output model is shown in \figref{fig:ParampInputOutput}. The nonlinear resonator is coupled with rate $\kappa$ to a transmission line, through which  the pump and signal fields propagate. Based on this model and the Hamiltonian in \eqref{eq:JPAHamiltonian} we obtain the following equation of motion for the intra-resonator field
\begin{equation}
\dot{A} = -i \omega_0 A - i K A^\dagger A A - \frac{\kappa + \gamma}{2} + \sqrt{\kappa} A_{\rm in}(t) + \sqrt{\gamma}b_{\rm in}(t).
\label{eq:EOMParamp}
\end{equation}
In addition to the coupling  to  transmission line modes $A_{\rm in}(t)$ with rate $\kappa$ we account for potential radiation loss mechanisms by introducing the coupling to modes $b_{\rm in}(t)$ with loss rate $\gamma$, compare \figref{fig:ParampInputOutput}(a). A boundary condition equivalent to
\begin{equation}
A_{\rm out}(t) = \sqrt{\kappa} A(t) - A_{\rm in}(t),
\label{eq:InOutBoundary}
\end{equation}
also holds for the loss modes. When operating the device as a parametric amplifier, the input field $A_{\rm in}$ is typically a sum of a strong coherent pump field  and an additional weak signal field. Since this signal carries at least the vacuum noise, it is treated as a quantum field. In this formalism  this particular situation is accounted for by decomposing each field mode into a sum of a classical part and a quantum part
\begin{eqnarray}
A_{\rm in}(t) &=& \left({a}_{\rm in}(t) + \alpha_{\rm in}\right)e^{-i \omega_p t},
\nonumber
\\ A_{\rm out}(t) &=& \left({a}_{\rm out}(t) + \alpha_{\rm out}\right) e^{-i \omega_p t},
\nonumber
\\
A(t) &=& \left({a}(t) + \alpha\right) e^{-i \omega_p t},
\label{eq:SignalPlusPump}
\end{eqnarray}
where  $\alpha$, $\alpha_{\rm in}$, $\alpha_{\rm out}$  represent the classical parts of the field which are associated with the pump, while ${a}$, $a_{\rm in}$, $a_{\rm out}$ account for the quantum signal fields. Since all $\alpha$'s are complex numbers the modes $a$ satisfy the same bosonic commutation relations as modes $A$ do. By multiplying  the field modes defined in \eqref{eq:SignalPlusPump} with the additional exponential factor $e^{-i \omega_{p} t}$, one works in a frame rotating at the pump frequency $\omega_p$. The strategy is to first solve the classical response for the pump field $\alpha$ exactly  and then linearize the equation of motion for the weak quantum field ${a}$ in the presence of the pump. Finally, we derive a scattering relation between input modes $a_{\rm in}$ and reflected modes $a_{\rm out}$.

The steady state solution  for the coherent pump field is determined by
\begin{equation}
\left((i(\tilde{\omega}_0-\omega_p) + \frac{\kappa +\gamma}{2}\right)\alpha + i K \alpha^2\alpha^* = \sqrt{\kappa}\alpha_{\rm in},
\label{eq:EOMCoherentComplex}
\end{equation}
which follows immediately by substituting \eqref{eq:SignalPlusPump} into \eqref{eq:EOMParamp} and collecting only the c-number terms. By multiplying both sides with their complex conjugate we get to the equation
\begin{eqnarray}
\frac{\kappa}{(\kappa + \gamma)^2}|\alpha_{\rm in}|^2 &=& \left(\left(\frac{\omega_p - \tilde{\omega}_0}{\kappa + \gamma}\right)^2 + \frac{1}{4}\right) |\alpha|^2
\nonumber
\\
&& \hspace{-10mm} - \frac{2 (\omega_p - \tilde{\omega}_0) K}{(\kappa + \gamma)^2} |\alpha|^4+ \left(\frac{K}{\kappa + \gamma}\right)^2 |\alpha|^6,
\label{eq:EOMC}
\end{eqnarray}
which determines the average number of pump photons $|\alpha|^2$ in the resonator. \eqref{eq:EOMC} reduces to
\begin{equation}
1=(\delta^2 + \frac{1}{4}) n - 2 \delta \xi n^2 + \xi^2 n^3,
\label{eq:EOMscaleInv}
\end{equation}
by defining the scale invariant quantities
 \begin{eqnarray}
 \delta &\equiv& \frac{\omega_p - \tilde{\omega}_0}{\kappa + \gamma},
 \nonumber
 \\
 \tilde{\alpha}_{\rm in} &\equiv& \frac{\sqrt{\kappa}\alpha_{\rm in}}{\kappa +\gamma},
 \nonumber
 \\
 \xi &\equiv &\frac{|\tilde{\alpha}_{\rm in}|^2  K}{\kappa +\gamma}
 \nonumber
 \\
 n &\equiv& \frac{|\alpha^2|}{|\tilde{\alpha}_{\rm in}|^2}.
 \label{eq:EOMreduced}
\end{eqnarray}
$\delta$ is the detuning between pump and resonator frequency  in units of the total resonator linewidth, $\tilde{\alpha}_{\rm in}$ is the dimensionless drive amplitude, and $\xi$ is  the product of drive power and nonlinearity, also expressed in dimensionless units. Finally, $n$ is the mean number of pump photons in the resonator relative to the incident pump power.  As an important consequence, we notice from \eqref{eq:EOMreduced} that only the product of drive power and nonlinearity determines the dynamics but not each quantity itself. Therefore, a small nonlinearity can at least in principle be compensated by increasing the drive power. Properties such as the gain-bandwidth product are therefore independent of the strength of the nonlinearity. Furthermore, the solutions of \eqref{eq:EOMscaleInv} for negative $\xi$ values are identical to those for positive $\xi$ up to a sign change in $\delta$. Since  $\xi$ is negative for the Josephson parametric amplifier, we focus on this particular case.

Equation (\ref{eq:EOMscaleInv}) is a cubic equation in $n$ and can therefore be solved analytically. We do not present the lengthy solutions here explicitly, but assume in the following that we have an explicit analytical expression for $n$ in terms of $\delta$ and $\xi$. In \figref{fig:ParampInputOutput}(b) we plot $n$ for various parameters $\xi$ as a function of $\delta$. At the critical value $\xi_{\rm crit} = -{1}/{\sqrt{27}}$ the derivative $\partial n /\partial \delta $ diverges and thus the response of the parametric amplifier becomes extremely sensitive to small changes.  For even stronger effective drive powers $\xi/\xi_{\rm crit} > 1$ the cubic \eqref{eq:EOMscaleInv} has three real solutions. The solutions for the high and low photon numbers are stable, while the intermediate one is unstable. The system bifurcates in this regime as mentioned earlier. The critical detuning below which the system becomes bistable is $\delta_{\rm crit} = -\sqrt{3}/2$. The critical point $(\xi_{\rm crit},\delta_{\rm crit})$ is the one at which both $\partial\delta/\partial n$ and $\partial^2\delta/\partial^2 n$ vanish. In scale invariant units the maximal value of $n$ is $4$, which is reached at the detuning $\delta=4 \xi$.

Experimentally, the device properties are characterized by measuring the complex reflection coefficient $\Gamma\equiv {\alpha_{\rm out}}/{\alpha_{\rm in}}$. Based on the input-output relation $\alpha_{\rm out} = \sqrt{\kappa} \alpha - \alpha_{\rm in}$ and \eqref{eq:EOMCoherentComplex} we evaluate this reflection coefficient as
\begin{eqnarray}
\Gamma = \frac{\kappa}{\kappa + \gamma}\; \frac{1}{\frac{1}{2} -i\delta  + i \xi n} - 1.
\label{eq:refCoeff}
\end{eqnarray}

In \figref{fig:ParampInputOutput}(c) we plot the absolute value of the reflection coefficient at $\xi=\xi_{\rm crit}$ for various loss rates $\gamma$. For vanishing losses $\gamma=0$ all the incident drive power is reflected from the device and $|\Gamma|=1$. Note that also in this case the resonance is clearly visible in the phase of the reflected signal (not shown here). When the loss rate $\gamma$ becomes similar to the external coupling rate $\kappa$ part of the radiation is dissipated into the loss modes. In the case of critical coupling $\gamma=\kappa$ all the coherent power is transmitted into the loss modes at resonance.  This is equivalent to the case of a  symmetrically coupled $\lambda/2$ resonator, for which the transmission coefficient is one at resonance \cite{Goppl2008}.
\subsection{Linearized response for weak (quantum) signal fields}
Under the assumption that the photon flux associated with the signal $\langle {a}^\dagger_{\rm in}{a}_{\rm in}\rangle$  is much smaller than the photon flux of the pump field $|\alpha_{\rm in}|^2$, we can drop terms such as $K a^\dagger a \alpha$, because they are small compared to the leading terms $K a^\dagger \alpha \alpha$ and $K a \alpha^* \alpha$. By neglecting these terms we obtain a linearized equation of motion for $a$ in the presence of the pump field. In order to preserve the validity of this approximation even for larger input signals, the amplitude $\alpha$ of the pump field needs to be increased. Experimentally, this can be achieved by reducing the strength of the nonlinearity $K$. Substituting \eqref{eq:SignalPlusPump} into \eqref{eq:EOMParamp} and keeping only terms which are linear in $a$ one finds
\begin{eqnarray}
\dot{a}(t) &=& i \left(\omega_p -\tilde{\omega}_0  - 2  K |\alpha|^2 + i \frac{\kappa+\gamma}{2}\right) a(t)
\nonumber
\\
&& - i K \alpha^2 a^\dagger(t) + \sqrt{\kappa}a_{\rm in}(t) +  \sqrt{\gamma}b_{\rm in}(t).
\label{eq:EOMFluctuations}
\end{eqnarray}
Since \eqref{eq:EOMFluctuations} is linear, we can solve it by decomposing all  modes into their Fourier components
\begin{eqnarray}
{a}(t) &\equiv&\sqrt{\frac{\kappa +\gamma}{2\pi}}\int_{-\infty}^\infty {\rm d}\Delta \,e^{- i \Delta (\kappa + \gamma) t} \, a_{\Delta}
\label{eq:aDefFreq}
\end{eqnarray}
and equivalently for ${a}_{{\rm in},\Delta}$ and ${b}_{{\rm in},\Delta}$. Note that  the detuning  $\Delta$ between signal frequencies and the pump frequency, is expressed here in units of the linewidth $\kappa + \gamma$. Substituting the Fourier decompositions into \eqref{eq:EOMFluctuations} and comparing the coefficients of different harmonics, results in
\begin{equation}
0 =  \left( i  (\delta  -   2\xi n + \Delta) - \frac{1}{2}\right) a_\Delta - i  \xi n e^{2 i  \phi}a^\dagger_{-\Delta} + \tilde{c}_{{\rm in},\Delta},
\label{eq:ResLinearized}
\end{equation}
where
$
\tilde{c}_{{\rm in},\Delta} \equiv ({\sqrt{\kappa}a_{{\rm in},\Delta}} + {\sqrt{\gamma}b_{{\rm in},\Delta}})/({\kappa +\gamma})
$
is the sum of all field modes incident on the resonator. Furthermore, in \eqref{eq:ResLinearized} $\phi$ is the phase of the intra-resonator pump field,  defined by $\alpha = |\alpha|e^{i \phi}$. The fact that  \eqref{eq:ResLinearized} couples  modes $a_{\Delta}$ and $a^\dagger_{-\Delta}$ can be interpreted as a wave mixing process. In order to express $a_\Delta$ in terms of the input fields $c_{\rm in,\Delta}$, \eqref{eq:ResLinearized} is rewritten as a matrix equation
\begin{widetext}
\begin{equation}
\begin{pmatrix}
  \tilde{c}_{{\rm in},\Delta} \\
 \tilde{c}^\dagger_{{\rm in},-\Delta} \\
  \end{pmatrix}
 =
 \begin{pmatrix}
  i \left(-\delta +   2\xi n  - \Delta\right)  +\frac{1}{2} &  i  \xi n e^{i 2 \phi}\\
 - i  \xi n e^{- i 2 \phi} &  i \left( \delta  - 2\xi n - \Delta \right)+\frac{1}{2}\\
  \end{pmatrix}
\begin{pmatrix}
  a_\Delta \\
 a^\dagger_{-\Delta}  \\
  \end{pmatrix}.
\end{equation}
By inverting the matrix on the right hand side, the quantum part of the intra-resonator field $a_\Delta$ is expressed in terms of the incoming field $\tilde{c}_{{\rm in},\Delta}$
\begin{equation}
a_{\Delta} = \frac{i \left( \delta  - 2\xi n - \Delta \right)+\frac{1}{2}}{(i\Delta - \lambda_-)(i\Delta - \lambda_+)} \tilde{c}_{{\rm in},\Delta}
+
\frac{ -i \xi n e^{2  i  \phi} }{(i\Delta - \lambda_-)(i\Delta - \lambda_+)} \tilde{c}^\dagger_{{\rm in},-\Delta}
\end{equation}
with
$\lambda_{\pm} = \frac{1}{2}\pm \sqrt{(\xi n)^2-(\delta - 2 \xi n)^2 }.$
Using \eqref{eq:InOutBoundary}, the final transformation between input and output modes is
\begin{subequations}
\begin{eqnarray}
\hspace{-10mm}
a_{\rm out,\Delta} &=& g_{S,\Delta} a_{\rm in,\Delta} + g_{I,\Delta} a^\dagger_{\rm in,-\Delta} + \sqrt{\frac{\gamma}{\kappa}} (g_{S,\Delta} + 1) b_{\rm in,\Delta} + \sqrt{\frac{\gamma}{\kappa}} g_{I,\Delta} b^\dagger_{\rm in,-\Delta}
\\
\label{eq:InOutParampA}
&\stackrel{\gamma/\kappa \rightarrow 0} {=}&
g_{S,\Delta} a_{\rm in,\Delta} + g_{I,\Delta} a^\dagger_{\rm in,-\Delta},
\label{eq:InOutParampB}
\end{eqnarray}
\end{subequations}
\end{widetext}
with
\begin{eqnarray}
g_{S,\Delta} &=& -1 + \frac{\kappa}{\kappa +\gamma}  \frac{i \left( \delta  - 2\xi n - \Delta \right)+\frac{1}{2}}{(i\Delta - \lambda_-)(i\Delta - \lambda_+)}
\label{eq:gain}
\end{eqnarray}
 {\rm and}\;\;
\begin{eqnarray}
g_{I,\Delta} = \frac{\kappa}{\kappa +\gamma} \frac{-i \xi n  e^{ 2 i  \phi} }{(i\Delta - \lambda_-)(i\Delta - \lambda_+)}.
\end{eqnarray}
\eqref{eq:InOutParampB} is the central result of this calculation. The output field at detuning $\Delta$ from the pump frequency is a sum of the input fields at frequencies $\Delta$ and $-\Delta$ multiplied with the signal gain factor $g_{S,\Delta}$ and the idler gain factor $g_{I,\Delta}$, respectively. The additional noise contributions introduced via the loss modes $b_{\rm in,\Delta}$ vanish in the limit $\gamma/\kappa\rightarrow 0$. In the ideal case $\gamma=0$, the coefficients  $g_{S,\Delta}$ and $g_{I,\Delta}$ satisfy the relation
\begin{equation}
G_\Delta \equiv |g_{S,\Delta}|^2 = |g_{I,\Delta}|^2 + 1
\end{equation}
and \eqref{eq:InOutParampB} is identical to a two-mode squeezing transformation \cite{Braunstein2005,Clerk2010} with gain $G_\Delta$. The two-mode squeezing transformation describes a linear amplifier in its minimal form (compare Ref.~\cite{Caves1982}), of which we discuss characteristic properties in the following section.
\subsection{Gain, bandwidth, noise and dynamic range}\label{sec:parampProps}
For simplicity we consider the case of no losses $\gamma=0$, for which the parametric amplifier response is described by
\eqref{eq:InOutParampB}. An incoming signal at detuning $\Delta$ is thus amplified by the power gain $G_\Delta=|g_{S,\Delta}|^2$ and mixed with the frequency components at the opposite detuning from the pump. Characteristic properties of the parametric amplifier, such as the maximal gain and the bandwidth, are thus encoded in the quantity $g_{S,\Delta}$ as a function of pump-resonator detuning $\delta$, effective drive strength $\xi$ and detuning between signal and pump $\Delta$.

\begin{figure}[b]
\centering
\includegraphics[scale=0.72]{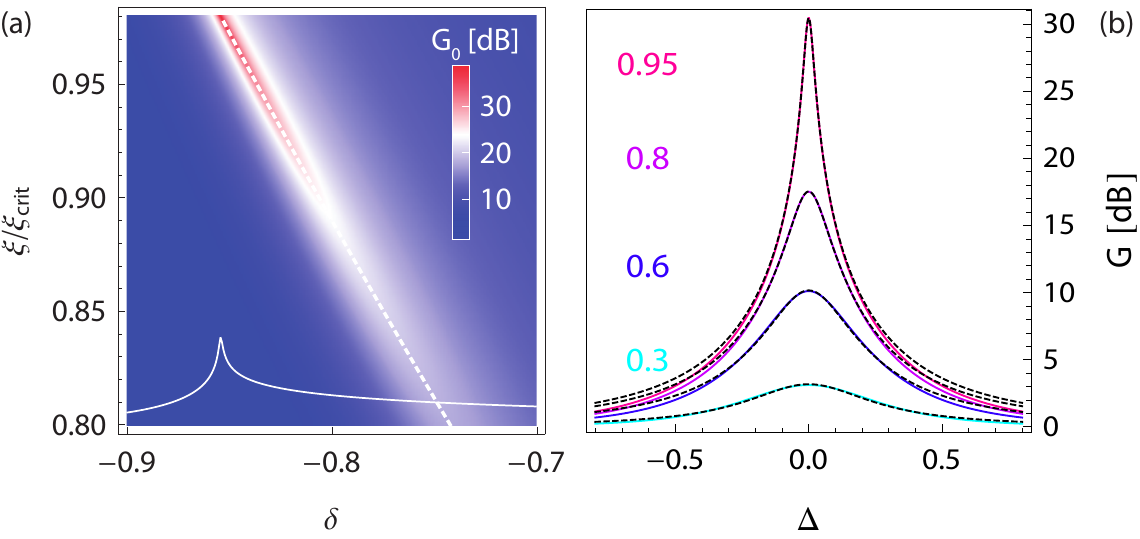}
\caption{(a)  Parametric amplifier gain  $G_\Delta=|g_{S,\Delta}|^2$ {\it vs.} pump tone detuning $\delta$ and drive strength $\xi$ at zero signal detuning $\Delta$ and for $\kappa=\gamma$. For increasing drive strength $\xi$ the detuning for maximum gain is indicated by the dashed white line. A cut through the data for the highest value $\xi=0.98 \xi_{\rm crit}$ is shown as the solid white line in the bottom part. (b) Gain as a function of signal detuning $\Delta$ for the indicated drives strengths $\xi_{}/\xi_{\rm crit}$ and optimal pump detuning. The exact gain curves (solid lines) are well approximated by Lorentzian lines (black dashed lines).
}
\label{fig:ParampTheory}
\end{figure}

In \figref{fig:ParampTheory}(a) we plot the gain $G_0$ for zero signal detuning $\Delta=0$ as a function of $\delta$ and $\xi$. We find that the maximal gain increases with increasing drive strength $\xi$ while the optimal value for $\delta$ at which this gain is reached, shifts approximately linearly with increasing $\xi$. The optimal values for $\delta$ are indicated as a dashed white line in \figref{fig:ParampTheory}(a). Mathematically, the gain diverges when $\xi$ approaches the critical value $\xi_{\rm crit}$. In practice, the gain is limited to finite values due to the breakdown of the stiff pump approximation \cite{Kamal2009} (see discussion below).

By changing the pump  parameters $\xi$ and $\delta$ we can adjust the gain $G_0$ to a desirable value, which is typically about $20\,$dB. Note that the gain can become smaller than one, in the presence of finite internal losses $\gamma>0$ . Once the pump parameters are fixed we characterize the bandwidth of the amplifier by analyzing the gain as a function of the signal detuning $\Delta$. In \figref{fig:ParampTheory}(b) we plot the gain as a function of $\Delta$ for the indicated values of $\xi/\xi_{\rm crit}$ and the corresponding optimal pump detunings $\delta$ (compare dashed white line in (a)). When the gain is increased, the band of amplification becomes narrower. This is quantitatively  expressed  by the gain bandwidth relation $\sqrt{G_0} B  \approx 1$, where $B$ is the detuning $\Delta$ for which the gain reaches half of its maximal value. Remember that $\Delta$ is defined in units of the resonator linewidth $\kappa + \gamma$, which means that the amplifier bandwidth equals approximately the resonator linewidth divided by the square root of the gain. The gain curves are well approximated by Lorentzian lines as indicated by the dashed black lines in \figref{fig:ParampTheory}(b). This Lorentzian approximation becomes  better with increasing gain.
The gain-bandwidth relation suggests that the amplifier bandwidth can be increased by lowering the total quality factor of the resonator. There are, however, several technical challenges to overcome when increasing the bandwidth. One of them is closely related to the dynamic range of the amplifier. In the derivation made in the previous sections we have assumed that the solution of the classical drive field is unaffected by the additional signal and quantum fluctuations at the input. This is known as the stiff pump approximation \cite{Kamal2009}, which assumes that the pump power  at the output is equal to the pump power at the input. This is of course an approximation, since the pump field provides the energy which is necessary for amplifying the input signal. The stiff pump approximation is valid as long as the pump power is significantly larger than the total output power of all amplified (quantum) signals and vacuum fluctuations \cite{Castellanos2008}. The minimum amount of energy transfer from the pump field into other modes is set by the amplification of vacuum noise within the band of amplification. According to \eqref{eq:InOutParampB} the integrated photon flux at the output of the JPA is equal to
\begin{equation}
P_{\rm out}\stackrel{\gamma=0}{=} \hbar \omega_p \kappa \int {\rm d}\Delta \langle a^\dagger_{\rm out,\Delta} a_{\rm out,\Delta}\rangle = \hbar \omega_p \kappa\int {\rm d}\Delta (G_\Delta-1),
\end{equation}
when only vacuum fluctuations are incident at the input. As an example, the realistic parameter configuration $\{\tilde{\omega}_0/2\pi,\kappa/2\pi,G_0\}=\{7\,{\rm GHz},100 \,{\rm MHz}, 20\,{\rm dB}\}$ corresponds to a power of amplified vacuum noise of about $-100\,$ dBm. If we want the pump power to be 20$\,$dB higher than this value, the Kerr nonlinearity $|K|/2\pi$ needs to be smaller than $\sim10\,$kHz, which is calculated using \eqref{eq:EOMreduced}. In \secref{sec:DynamicRange} we discuss how this nonlinearity can be decreased by making use of multiple SQUIDs connected in series.

When operating the JPA, we also have to understand its behavior in terms of added noise.  In the ideal case with zero loss rate ($\gamma=0$), the input-output relation  of the parametric amplifier in \eqref{eq:InOutParampB} has the minimal form of a scattering mode amplifier \cite{Clerk2010}. The amplification process reaches the vacuum limit as long as the input modes are cooled into the vacuum. In practice, however, the device may have finite loss $\gamma$ which increases the effectively added noise by a factor of  $(\kappa + \gamma)/\kappa$. This is due to the additional amplified noise, which originates from the modes $b_{\rm in,\Delta}$ and contributes to the output field $a_{\rm out,\Delta}$ (compare \eqref{eq:InOutParampA}).  Another potential source of noise is  related to the stability of the resonance frequency of the parametric amplifier. Magnetic flux noise in the SQUID loop may lead to a fluctuating resonance frequency and thus a fluctuating effective gain.
\section{Effective system parameters from distributed circuit model}
\label{sec:tunableRes}
In the previous section we have analyzed the model of a nonlinear resonator with resonance frequency $\tilde{\omega}_0$, Kerr nonlinearity $K$ and decay rate $\kappa$. Here, we explicitly derive this effective Hamiltonian from the full circuit model of a $\lambda/4$ - transmission line resonator, which is terminated by a SQUID loop at the short-circuited end and coupled capacitively to a transmission line, see \figref{fig:ParampSchematic}(a). These calculations allow us to determine $\tilde{\omega}_0,K,\kappa$ from the distributed circuit parameters and give insight into potential limitations of the effective model. We also compare the obtained parameter relations with those of a lumped element parametric amplifier.
\subsection{Resonator mode structure in the linear regime}
\begin{figure}[t]
\centering
\includegraphics[scale=1.]{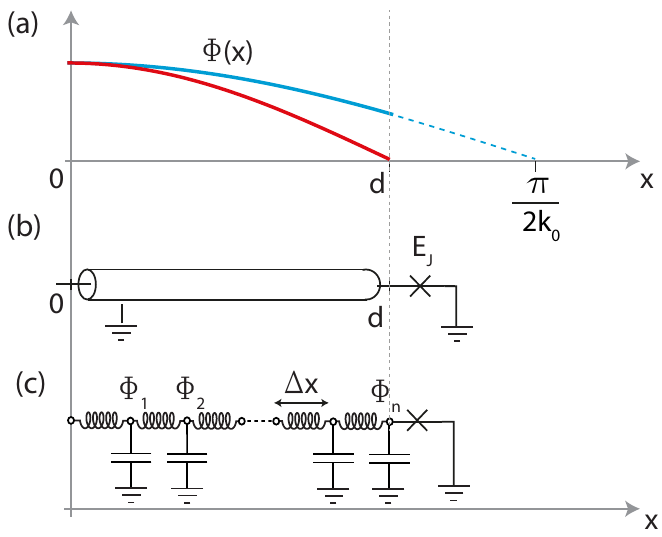}
\caption{(a) Distribution of the magnetic flux field $\Phi(x)$ along the $\lambda/4$-resonator for the fundamental resonator mode $j=0$, without (red) and with (blue) a Jospehson  junction. The additional Josephson inductance changes the boundary condition such that neither the current nor the voltage is zero at position $x=d$. The resulting increase in the effective wavelength $\pi/(2 k_0)$ is indicated by the dashed blue line. (b) Transmission line resonator of length $d$ with a Josephson junction at the grounded end. (c) Lumped element representation with indicated discretized magnetic flux field $\Phi_j$ as used in \eqref{eq:LTot}.
}
\label{fig:parampSpec1}
\end{figure}

In order to find the normal mode structure of the system, we first neglect its capacitive coupling  to the transmission line as indicated in \figref{fig:parampSpec1}. The derivation is similar to the one in Ref.~\cite{Wallquist2006}. Dissipation effects due the environment are discussed in \secref{sec:lowQCorrection}.

The total Lagrangian of the system in the magnetic flux field $\Phi(x)$ has a transmission line part and a term which describes the SQUID at position $x=d$ (\figref{fig:parampSpec1}).
\begin{eqnarray}
\mathcal{L} &=& \int_{0}^d {\rm d}x \Big{\{}  \frac{c}{2} \left(\partial_t \Phi(x)\right)^2  - \frac{1}{2l} \left(\partial_x \Phi(x)\right)^2 \Big{\}}
\nonumber
\\
& &+ E_{J} \cos\left(\frac{\Phi(d)}{\varphi_0}\right)
\label{eq:LTot}
\end{eqnarray}
with the reduced flux quantum $\varphi_0=\hbar/2e$.
Since we work in a limit in which the plasma frequency of the SQUID is much larger than the resonance frequencies of interest, we can  neglect the self-capacitance of the SQUID. The SQUID is furthermore described as a single junction with tunable effective Josephson energy $E_{J}$.

\begin{figure}[t]
\centering
\includegraphics[scale=1.]{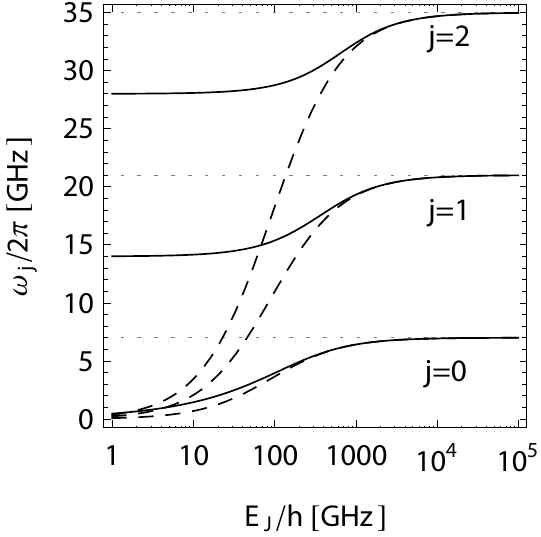}
\caption{Resonance frequencies of the first three modes as a function of Josephson energy. The solid line results from the exact numerical solution of \eqref{eq:dispRel} while the dashed line shows the linearized solution in \eqref{eq:dispRelApprox}. The bare resonance frequency is chosen to be $7\,$GHz and the impedance of the transmission line resonator 50$\,\Omega$.
}
\label{fig:parampSpec2}
\end{figure}

We first investigate the linear regime of the system, in which the cosine potential of the SQUID is approximated as a quadratic potential.
\begin{equation}
E_{\rm J} \cos\left(\frac{\Phi(d)}{\varphi_0}\right) \approx {\rm const} - \frac{1}{2}\left(\frac{\Phi(d)}{\varphi_0}\right)^2
\end{equation}
Due to the spatial derivative in the Lagrangian in \eqref{eq:LTot} all local fields in the chain are coupled to their next neighbors and the the normal mode structure is found by solving the Euler-Lagrange equation $ \partial_t ({\delta \mathcal{L}}/{\delta \dot{\Phi}}) -  {\delta \mathcal{L}}/{\delta {\Phi}}= 0$ of the transmission line resonator. This   results in the wave equation
\begin{eqnarray}
v^2 \partial_x^2 \Phi(x) - \partial_t^2 \Phi(x) =0,
\label{eq:waveEq}
\end{eqnarray}
with the phase velocity $v=1/\sqrt{cl}$, of which the general solution can be written as a sum of normal modes
\begin{eqnarray}
\Phi(x) = \sum_{j=0}^\infty \phi_j \cos(k_j x).
\label{eq:decomposition}
\end{eqnarray}
The valid wavevectors $k_j$ are determined by the boundary conditions at the two ends of the $\lambda/4$ transmission line.
The open end at $x=0$  requires that the current $\partial_x \Phi(x)/l$ vanishes, which is implicitly satisfied by choosing the cosine ansatz in \eqref{eq:decomposition}. On the shorted end the boundary condition is modified by the presence of the Josephson junction. In order to determine this boundary condition, we evaluate the Euler-Lagrange equation at position $x=d$. For this purpose it is  convenient to write the Lagrangian in a discretized form, see \figref{fig:parampSpec1}(c) and compare Ref.~\cite{Wallquist2006}:
\begin{eqnarray}
\mathcal{L} &=& \lim_{n \rightarrow \infty} \sum_{j=1}^n {\Delta}x \Big{\{}  \frac{c}{2} (\partial_t \Phi_j)^2  - \frac{1}{l} \frac{(\Phi_j
-\Phi_{j-1})^2}{{\Delta}x^2} \Big{\}}
\nonumber
\\
& &- \frac{1}{2}E_{\rm J} \left(\frac{\Phi_n}{\varphi_0}\right)^2
\label{eq:LTotDisc}
\end{eqnarray}
where $\Phi_n = \Phi(x=d)$ and $\Delta x=d/n$. Evaluating $\partial_t ({\partial \mathcal{L}}/{\partial \dot{\Phi}_n}) -  {\partial \mathcal{L}}/{\partial {\Phi_n}}= 0$ leads to the equation
\begin{equation}
\frac{1}{l}\partial_x\Phi(d) +E_{\rm J} \frac{\Phi(d)}{\varphi_0^2}=0.
\label{eq:bound}
\end{equation}
Substituting  the ansatz (\ref{eq:decomposition}) into \eqref{eq:bound} and comparing the resulting coefficients of  the independent variables $\phi_j$, results in the transcendental equation
\begin{eqnarray}
k_j d \tan(k_j d) =  l d \frac{E_{\rm J}}{\varphi_0^2} \equiv \frac{l d}{ L_{\rm J}}.
\label{eq:dispRel}
\end{eqnarray}
Here, we have defined the Josephson inductance $L_{J} = {\varphi_0^2}/{E_{\rm J}}$. The infinite set of solutions $k_j$ of this equation determines the normal modes structure of the system in the linear regime. In the limit in which the SQUID inductance $L_J$ vanishes, \eqref{eq:dispRel} is solved by the poles of $\tan(k_j d)$, and we recover the normal modes of the $\lambda/4$ resonator
\begin{equation}
k^{(0)}_j d = \frac{\pi}{2}(1+ 2 j)\;\;{\rm with}\;\;j \in \{0,1,2,3,...\}.
\end{equation}
As a first order correction to this result in the limit of $L_{\rm J}/l d \ll 1$, we expand \eqref{eq:dispRel} to first order in $(k^{(0)}_j-k_j)d$ and find $k_j L_{\rm J}/ld = (k^{(0)}-k_j)$ or equivalently
\begin{eqnarray}
k_j \approx \frac{k_j^{(0)}}{1 + {L_{\rm J}}/{ld}}.
\label{eq:dispRelApprox}
\end{eqnarray}
For the fundamental mode with $j=0$ this linearized approximation is typically accurate even for inductance ratios up to $L_{\rm J}/l d\approx 0.5$, whereas for the higher harmonic modes the linearized equation breaks down for much smaller values of $L_{\rm J}/ld$. A comparison between the exact solution based on \eqref{eq:dispRel} and the approximate solution in  \eqref{eq:dispRelApprox} is shown in \figref{fig:parampSpec2} for the first three resonant modes. When higher harmonics are expected to be relevant one should solve \eqref{eq:dispRel} numerically in order to determine the exact wave numbers $k_j$.
\subsection{Kerr nonlinear terms and effective Hamiltonian}
Using the normal mode decomposition in \eqref{eq:decomposition} we reexpress the  Lagrangian in \eqref{eq:LTot} as a sum of oscillators which are only coupled via the boundary condition imposed by the SQUID. For the purposes of parametric amplification the phase drop across the junction is desired to be small, $\Phi_n/\varphi_0 < 1$, i.e. the current flowing through the Josephson junction is small compared to its critical current. We can therefore expand the SQUID cosine potential and take into account only the first non-quadratic correction
\begin{equation}
E_{\rm J} \cos\left(\frac{\Phi_n}{\varphi_0}\right)^2= {\rm const}  - \frac{1}{2}E_{\rm J}\left(\frac{\Phi_n}{\varphi_0}\right)^2 + \frac{1}{24} E_{\rm J}\left(\frac{\Phi_n}{\varphi_0}\right)^4 + ...
\label{eq:CosExpand}
\end{equation}
In \secref{sec:DynamicRange} we discuss under which circumstances such an approximation may break down. Substituting the normal mode decomposition \eqref{eq:decomposition} into  the Taylor expansion of the Lagrangian results in
\begin{equation}
\mathcal{L} =  \frac{1}{2}\sum_{i=1}^\infty  \big{\{}  \dot{\phi}_i{C}_{i}\dot{\phi}_i  - \phi_i{L}^{-1}_{i} \phi_i \big{\}}
+ \sum_{j,i,k,l=1}^\infty  N_{ijkl} \phi_i \phi_j \phi_k \phi_l
\label{eq:LTotNormal}
\end{equation}
with the effective capacitances and inductances \cite{Wallquist2006}
\begin{eqnarray}
{C}_{i} &\stackrel{}{=}&c \int_{0}^d {\rm d}x \cos^2(k_i x) = \frac{cd}{2}\left(1+\frac{\sin(2 k_i d)}{2 k_i d}\right),
\nonumber
\\
{L}^{-1}_{i} &\stackrel{}{=}&   L_{\rm J}^{-1}  \cos^2(k_i d) + \frac{ k_i^2}{l} \int_{0}^d {\rm d}x \sin^2(k_i x)
\nonumber
\\
&\stackrel{{\rm Eq.}~(\ref{eq:dispRel})}{=}& \frac{ (k_i d)^2}{2 l d} \left(1+\frac{\sin(2 k_i d)}{2 k_i d}\right),
\label{eq:CLKDef}
\end{eqnarray}
and the nonlinearity coefficients
\begin{eqnarray}
N_{ijkl} &=& \frac{1}{24}E_{\rm J}\varphi_0^{-4} \prod_{m \in\{i,j,k,l\}}\cos(k_m d)\,.
\end{eqnarray}
As expected the linear part of the Lagrangian is diagonal in the normal mode basis. It describes a set of uncoupled $LC$ oscillators for which the effective resonance frequencies coincide with the product of phase velocity and wave vector $\omega_j = k_j v =1/\sqrt{{L}_{j} { C}_{j}}$.

Based on the Lagrange function~(\ref{eq:LTotNormal}) we derive the Hamiltonian by introducing the conjugate charge variables
$
q_i = {\delta \mathcal{L}}/{\delta \dot{\phi_i}}
= C_{i} \dot{\phi}_i.
$
Performing a Legendre transformation and taking only self-interactions and two-mode interactions into account, results in the Hamiltonian
\begin{eqnarray}
\mathcal{H} &=&  \frac{1}{2}\sum_{i=1}^\infty  \big{\{}  q_i{C}^{-1}_{i}q_i  + \phi_i{ L}^{-1}_{i} \phi_i \big{\}}
\nonumber
\\
&& \hspace{-0 mm}- 3\sum_{j\neq i}^\infty  N_{iijj} \phi_i^2 \phi_j^2
- \sum_{i}^\infty  N_{iiii} \phi_i^4.
\label{eq:HamiltonianFull}
\end{eqnarray}
In a quantum regime $q_i$ and $\phi_i$ are operators which satisfy the commutation relation $[\phi_j,q_k]=\delta_{kj}\hbar/i$ and it is convenient to write the Hamiltonian in terms of normal mode annihilation and creation operators \cite{Girvin2011}
\begin{eqnarray}
\phi_{j} &=&  i \phi_{{\rm zpf},j}(  a_j^\dagger -a_j)
\;\;,\;\;
q_{j} = q_{{\rm zpf},j}( a_j + a_j^\dagger)
\label{eq:modeOperators}
\end{eqnarray}
with
$
q_{{\rm zpf},i} = \sqrt{{\hbar \omega_i {C}_{i}}/{2 }}$ and
$
\phi_{{\rm zpf},i} = \sqrt{{\hbar }/{ 2 \omega_i {C}_{i}}}.
$
The abbreviation zpf stands for zero point fluctuations. Performing a rotating wave approximation (i.e.~removing all terms with an unequal number of creation and annihilation operators), and neglecting the small photon number independent frequency shifts due to the nonlinear terms (i.e. Lamb shifts) we arrive at
\begin{eqnarray}
\mathcal{H} =  \sum_{i=1}^\infty  \hbar \omega_{i}a_i^\dagger a_i + \hbar \frac{K_{ii}}{2}  a_i^\dagger a_i^\dagger a_i  a_i
+ \sum_{j\neq i}^\infty  \hbar K_{ij} a_i^\dagger a_i a_j^\dagger a_j .
\label{eq:HamiltonianFinal}
\end{eqnarray}
with
\begin{eqnarray}
K_{ij} &=& -\frac{ E_{\rm J}}{2 \hbar}\left(\frac{\phi_{\rm zpf}}{\varphi_0}\right)^4 \cos^2(k_i d)\cos^2(k_j d).
\label{eq:Kerr}
\end{eqnarray}
The quantity $K=K_{00}$ is the Kerr nonlinearity of the fundamental mode, which is used for the parametric amplification process. The terms proportional to $K_{ij}$ with unequal $i\neq j$ are cross Kerr interaction terms which  couple different modes to each other. Such an interaction can for example be used for counting the number of photons in one mode by probing another one with a coherent field \cite{Imoto1985,Sanders1989,Santamore2004,Buks2006,Helmer2009b,Johnson2010,Suchoi2010,Kirchmair2013}, similarly to a dispersive qubit measurement. Note that the values resulting from \eqref{eq:Kerr} are divided by the square of the number of SQUIDs, if an array is used instead of a single SQUID, as discussed in the following section.
\subsection{Decay rate and resonance frequency correction for low $Q$ resonators}
\label{sec:lowQCorrection}
Since the parametric amplifier bandwidth is proportional to the decay rate $\kappa$, typical devices are  designed to have a low external quality factor, which is achieved by increasing the coupling capacitance $C_\kappa$ between transmission line and resonator (\figref{fig:ParampSchematic}). The coupling of an oscillator to the environment shifts its resonance frequency $\omega_j \rightarrow \tilde{\omega}_j$ \cite{Goppl2008}, which can be significant if the coupling rate is large. When designing parametric amplifier devices,  it is therefore necessary to take these shifts into account. Based on the effective inductance and capacitances calculated in \eqref{eq:CLKDef} we find
\begin{equation}
\tilde{\omega}_j^2 \approx \frac{\omega_j^2}{1+ {C_\kappa}/{C_j}} = \frac{1}{(C_j + C_\kappa)L_j} \,\,{\rm and}\,\, \kappa_j \approx \frac{\tilde{\omega}_j^2 C_\kappa^2 R}{C_\kappa +  C_j}
\label{eq:lowQShift}
\end{equation}
for resonance frequency and decay rate of the $j$'th mode of the parametric amplifier device. The external quality factor is given by $Q_j\equiv\tilde{\omega}_j/\kappa_j$.
\subsection{Lumped element JPA}
As already mentioned in the introduction, a JPA can also be realized as a lumped element resonator by shunting a SQUID with a large  capacitance $C_J$ \cite{Vijay2011,Hatridge2011}. In this case the resonator is described by the transmon Hamiltonian \cite{Koch2007}, which in the deep transmon limit $E_J\gg E_C\equiv e^2/2C_J$ takes the form of \eqref{eq:JPAHamiltonian} with anharmonicity $K\approx E_C/\hbar$ and resonance frequency $\tilde{\omega}_0 \approx 1/\sqrt{L_J C_J}$. Also for this type of resonators the coupling rate $\kappa$ to the transmission line can be designed independently of $E_J$ and $E_C$ by designing an appropriate capacitive network. Similarly as for the transmission line JPA, the description in terms of the effective Hamiltonian \eqref{eq:JPAHamiltonian} is based on the assumption that for relevant resonator fields the phase drop across the Josephson junctions is small (compare \eqref{eq:CosExpand}). In the following section we study the validity of this approximation when the resonator is driven close to the bifurcation point where we expect parametric amplification to occur and analyze its implications for realizing a parametric amplifier with large bandwidth and dynamic range.
\section{Bandwidth and dynamic range constraints}
\label{sec:DynamicRange}
\subsection{Validity of the quartic approximation}
For deriving the Hamiltonian in \eqref{eq:JPAHamiltonian}, or more generally \eqref{eq:HamiltonianFinal}, we have expanded the SQUID cosine potential to quartic order in the dimensionless flux variable $\Phi_n/\varphi_0$, where $\Phi_n\equiv\Phi(x=d)$ is the phase drop across the SQUID. To guarantee that this approximation holds when we operate the device in the parametric amplification regime, we have to make sure that $\Phi_n/\varphi_0$ is small even close to the bifurcation point. This is equivalent to evaluating if the current flowing through the SQUID at corresponding drive powers is  small compared to the critical current.

To characterize the validity of the low order expansion of the cosine potential, we define the maximal coherent field inside the resonator $\alpha_{\rm max}$  as the one for which $\Phi_n \stackrel{!}{=}\varphi_0$. This is the coherent amplitude, at which the current flowing through the SQUID equals its critical current. According to \eqref{eq:modeOperators} and \eqref{eq:decomposition} a coherent field $\alpha$ in mode $j$ leads to a maximal amplitude of $\Phi_n\stackrel{}{=}\phi_{{\rm zpf},j}2\alpha \cos(k_j d)$ across the tunnel junctions, based on which we define the critical amplitude as
\begin{equation}
\alpha_{{\rm max},j} = \frac{\varphi_0}{\phi_{{\rm zpf},j}} \frac{1}{{2} \cos(k_j d) }
= \frac{\varphi_0}{\phi_{{\rm zpf},j}} \frac{l d}{{2} L_J  k_j d \sin(k_j d) }.
\end{equation}
The low order expansion of the cosine potential is only valid if the field inside the resonator $\alpha$ is  much smaller than this maximal amplitude $\alpha < \alpha_{\rm max}$.
\begin{figure}[t]
\centering
\includegraphics[scale=1.]{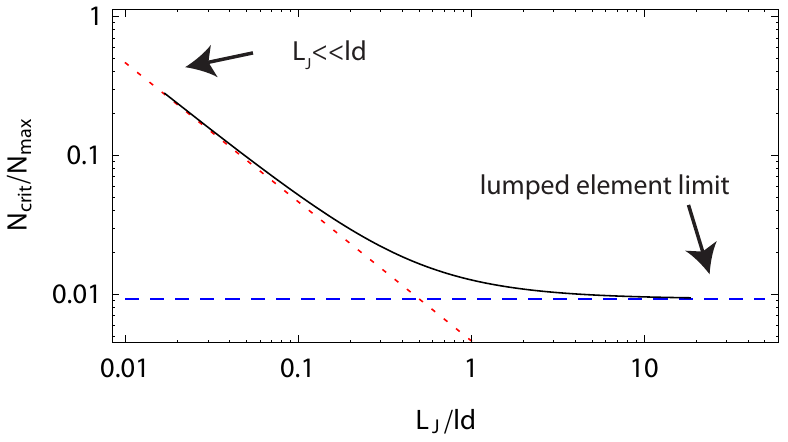}
\caption{Ratio $N_{\rm crit}/N_{\rm max}$ according to \eqref{eq:NCritRatio} (black), in the lumped element limit (dashed blue) and in the limit of small participation ratio $L_J\ll l d$ (dotted red) for the parameters $\{\tilde{\omega}_{0}/2\pi,Q\} = \{7\,{\rm GHz}, 1000\}$}
\label{fig:NCritNMax}
\end{figure}
In \secref{sec:parampInOut} we have found that the photon number in a resonator mode at the bifurcation point is $N_{\rm crit} = (\kappa+\gamma)/\sqrt{3}K$. The ratio between $N_{\rm crit}$ and the maximal coherent photon number $N_{{\rm max},j} \equiv |\alpha_{{\rm max},j}|^2 $, which we would like to keep small, is given by
\begin{eqnarray}
\frac{N_{\rm crit}}{N_{{\rm max},j}}&=& \frac{8 \kappa}{\sqrt{3}}\frac{\hbar}{E_{\rm J} \cos^2(k_j d)} \big(\frac{\varphi_0}{\phi_{\rm zpf}}\big)^2
\label{eq:NCritRatio}
\end{eqnarray}
with the two interesting limits
\begin{equation}
\frac{N_{\rm crit}}{N_{{\rm max},j}}
 =  \begin{cases} \frac{8}{\sqrt{3}} Q^{-1} \frac{l d}{L_{\rm J}}, &
  {\rm for}\,L_J\ll l d\\  \frac{16}{\sqrt{3}} Q^{-1}, & {\rm for\,lumped\,element\,JPA} \end{cases}.
\end{equation}
\begin{figure*}[t]
\centering
\includegraphics[scale=1.]{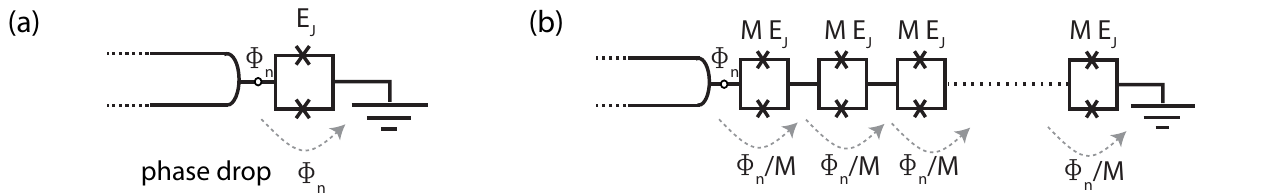}
\caption{(a) The phase drop across a single SQUID junction is proportional to the node flux $\Phi_n$ (indicated by the circle) at the end of the transmission line. (b) If we replace the single junction by an serial array of $M$ junctions with $M$ times larger Josephson energy, the phase drop across each junction is by a factor of $M$ smaller while the total effective Josephson inductance stays the same.}
\label{fig:SQUIDArray}
\end{figure*}

This is an important result, which sets clear constraints on both the maximally achievable bandwidth and the dynamic range of the JPA. Increasing the bandwidth, i.e. decreasing $Q$, leads to an increase of ${N_{\rm crit}}/{N_{{\rm max},j}}$. On the other hand a larger bandwidth requires a larger pump field in order to provide the necessary power to amplify (at least) the vacuum noise. Larger pump fields can only be achieved by decreasing the Kerr nonlinearity $K$, which requires a decrease in $L_J$. This leads, on the other hand, to a larger ratio $N_{\rm crit}/N_{\rm max}$  as illustrated in \figref{fig:NCritNMax}.

Interestingly, we find that in the lumped element case the Josephson inductance $L_J$, and with it the Kerr nonlinearity $K$, can in principle be made smaller without affecting $N_{\rm crit}/N_{\rm max}$. However, in practice a  small Josephson inductance has to be compensated by a large lumped element capacitor to retain the desired resonance frequency, which is challenging to realize without introducing additional parasitic geometric inductances. It therefore seems difficult to build a parametric amplifier with large bandwidth and high dynamic range at the same time using a single SQUID only. In the following we show how one can keep  $N_{\rm crit}/N_{\rm max}$ constant while decreasing the nonlinearity and thus increasing the dynamic range of the amplifier, by replacing the single SQUID with a serial array of $M$ SQUIDs of $M$-times larger Josephson energy per SQUID (\figref{fig:SQUIDArray}).
\subsection{Josephson junction arrays}
For simplicity we assume that all SQUIDs in the array have the same effective Josephson energy $M E_J$. Since the spatial extent of the junction is still small compared to typical resonance wavelengths, we can treat the array as a lumped element. To derive the nonlinearity of the oscillator for this situation we investigate how the different terms in the Lagrangian scale with $M$.

Assuming that the phase drop from the flux node at the end of the transmission line resonator to the ground is  homogeneously distributed over the array,  we have the same phase drop $\Phi_n/M$ across each SQUID, see \figref{fig:SQUIDArray}. As a result the quadratic term in the Lagrangian scales as
\begin{eqnarray}
\frac{E_{\rm J}}{2} \Phi^2_n \stackrel{1 \rightarrow M}{\longrightarrow} \sum_{i=1}^M \frac{M E_{\rm J}}{2} \left(\frac{\Phi_n}{M}\right)^2 = \frac{E_{\rm J}}{2} \Phi^2_n
\end{eqnarray}
and thus remains constant. This agrees with our expectation, since the total linear Josephson inductance has not been changed. However, the quartic term scales like
\begin{eqnarray}
\frac{E_{\rm J}}{24} \Phi^4_n \stackrel{1 \rightarrow M}{\longrightarrow} \sum_{i=1}^M \frac{M E_{\rm J}}{24} \left(\frac{\Phi_n}{M}\right)^4 = \frac{1}{M^2}\frac{E_{\rm J}}{24} \Phi^4_n,
\end{eqnarray}
which leads to a quadratic decrease in the effective Kerr nonlinearity $K \rightarrow K/M^2$ and thus a quadratic increase in $N_{\rm crit}\propto M^2$. Furthermore, the maximal photon number also scales as $ N_{{\rm max}} \propto M^2$ since the critical current of each junction is larger by a factor of $M$. In other words, the ratio $N_{\rm crit}/N_{\rm max}$  only depends on the total Josephson inductance whereas the bifurcation power increases quadratically in $M$. We thus conclude that the dynamic range of a JPA can be increased without affecting the amplifier bandwidth, by using an array of SQUIDs instead of a single SQUID. This conclusion is valid for both the transmission line JPA and the lumped element JPA.

In practice, the Josephson energies in the array are not all equal due to inhomogeneous coupling to the external magnetic flux and scatter in the critical current of Josephson junctions due to unavoidable variations in fabrication. A quantitative analysis of the influence of such variations of Josephson energies on the parametric amplifier characteristics could be an interesting task for future studies. This would help to quantify limitations in the accessible tuning range of the parametric amplifier and a realistic understanding of the breakdown of the low order expansion of the cosine potential. For such an approach the methods used in Ref.~\cite{Ferguson2013} could turn out to be useful.
\section{Conclusion}
In summary, we have presented a detailed analysis of Josephson junction based parametric amplifiers, including a discussion of bandwidth, noise and dynamic range. By establishing relations between basic JPA properties and designable circuit parameters we have been able to derive two simple design strategies to achieve optimized JPA performance. On the one hand the contribution of the Josephson inductance to the total effective inductance of the resonant circuit has to be chosen sufficiently large. On the other hand the use of SQUID arrays instead of single SQUIDs provides the possibility to enhance the strength of the pump field close at the bifuraction point and with it the dynamic range of the JPA.
\bibliographystyle{../myapsrev}

\end{document}